\documentclass[a4paper,10pt]{article}
\usepackage{latexsym}
\usepackage{amssymb}
\usepackage{graphicx}
\usepackage{color}

\begin{document}

\rightline{\large IFIC-XXX}

\vspace*{2.0cm}

\begin{center}
{\LARGE \textbf{
$H\to \ell\ell'$ in the Simplest Little Higgs Model
\\[2cm]}}

{\large \textbf{Andrea Lami$^{1}$ and Pablo Roig$^{2}$}\\[1.2 cm]}
$\ ^{1}$\textit{Instituto de F\'{\i}sica Corpuscular, Universitat de Val\`encia - CSIC, Apt. Correus 22085, E-46071 Val\`encia, Spain}\\[0.4cm]
$\ ^{2}$\textit{Departamento de F\'{\i}sica, Centro de Investigaci\'on y de Estudios Avanzados del Instituto Polit\'ecnico
Nacional, Apartado Postal 14-740, 07000 M\'exico D.F., M\'exico}
\end{center}

\vspace*{2.0cm}
\begin{abstract}
Little Higgs Models are promising constructs to solve the hierarchy problem affecting the Higgs boson mass for generic new physics. However, their preservation of lepton universality 
forbids them to account for the $H\to\tau\mu$ CMS hint and at the same time respect (as they do) the severe limits on $H\to\mu e$ inherited from the non-observation of $\mu\to e\gamma$. 
We compute the predictions of the Simplest Little Higgs Model for the $H\to \ell\ell'$ decays and conclude that the measurement of any of these decays at LHC (even with a much smaller 
rate than currently hinted) would, under reasonable assumptions, disfavor this model. This result is consistent with our earlier observation of very suppressed lepton flavor violating 
semileptonic tau decays within this model.

\end{abstract}
\vspace*{2.0cm}

\section{Introduction}
The discovery of a scalar boson at the LHC experiments ATLAS \cite{Aad:2012tfa} and CMS \cite{Chatrchyan:2012xdj} with properties remarkably close to the one proposed as agent of 
electroweak symmetry breaking in 1964 \cite{BEH} has explained the origin of elementary particle masses (with maybe neutrinos aside) within the Standard Model (SM). Despite the 
extraordinary success of this theory, there remain some unanswered questions within the SM, such as the origin of dark matter or the explanation of the baryon asymmetry of the universe. 
These and other puzzles motivate the scrutiny of the SM predictions against measurements in the search for new physics.\\
One way to do this is to take advantage of observables where both sharp predictions and accurate measurements are possible, in such a way that tiny deviations between them can become 
statistically significant and hint at the new dynamics. An alternative is to search for forbidden processes in the SM, where an observation must be due to new phenomena. Lepton flavor 
violating (LFV) processes are practically forbidden in the minimal extension of the SM with right-handed neutrinos, where the large suppression given by the ratio between the neutrino 
masses and electroweak scale makes them undetectable \cite{Petcov:1976ff}.\\
Although currently well compatible with the SM, the uncertainties associated to Higgs physics still leave room for sizable non-standard decays, such as invisible modes \cite{invisibles} 
or LFV decays. In the latter case, which we consider in this letter, direct searches for $H\to\tau\mu$ decays have been performed by the ATLAS and CMS Collaborations. The CMS result 
\cite{Khachatryan:2015kon}, $BR(H\to\tau\mu)=\left(0.84^{+0.39}_{-0.37}\right)\%$ ($<1.51\%$ at $95 \%$ CL) caused a big excitement in the community, although ATLAS \cite{Aad:2015gha} did not 
confirm it a few months later, setting the bound $BR(H\to\tau\mu)<1.85\%$ at the same confidence level \footnote{Here, and throughout the letter, the notation $BR(H\to\tau\mu)$ implies 
$BR(H\to\tau^-\mu^+)+BR(H\to\tau^+\mu^-)$.} \footnote{CMS has just presented their update including 13 TeV data \cite{Cepeda}, $BR(H\to\tau\mu)=\left(-0.76\pm0.81\right)\%$, 
which lowers the combined significance of the hint at the two sigma level.}.\\
The possibility that this branching fraction could be measurable was put forward much early in diverse SM extensions \cite{earlypapers}, typically including several scalar doublets 
\cite{Bjorken:1977vt}, and reanalyzed including the limits on $L\to\ell\gamma$ decays and other low-energy processes in forthcoming years \cite{LowE}. The CMS hint was the object of 
several devoted studies which could find regions of their parameter space explaining the anomaly \cite{explanations}. However, some other analyses either found difficult to produce such a 
strong signal in well-motivated extensions of the SM or pinpointed tensions with other precision measurements \cite{noexplanations, Yang:2016hrh}.\\
It is well known that the hierarchy problem in the Higgs mass arises whenever heavy new particles are introduced without a mechanism devised to cancel or effectively suppress the corresponding  
radiative corrections. One such mechanism assumes that the Higgs boson is a pseudo-Nambu-Goldstone boson of a spontaneously broken (larger than in the SM) electroweak symmetry group, in 
the so-called Little Higgs models \cite{LH}. Simplest Little Higgs (SLH) model \cite{SLH} is a concrete realization of this idea that we employed in our study of LFV semileptonic tau 
decays recently \cite{Lami:2016vrs}. Our main conclusion was that the observation of LFV semileptonic tau decays at Belle-II \cite{Wang:2015kmm} would (using reasonable approximations) 
rule out this model, as it predicts 
decay rates four to five orders of magnitude smaller than current bounds. According to LHC results, $H\to \ell\ell'$ decays are allowed at a rate of about $1\%$. However, the severe 
limits on $\mu\to e\gamma$ constrain $BR(H\to \mu e)\lesssim10^{-8}$ \cite{LowE} and a strong suppression on $BR(H\to \tau \mu)$ is also expected from the 
$\tau\to \mu\gamma$ bounds \cite{Yang:2016hrh}. Since the SLH model does not violate lepton universality (LU), confirmation of the CMS hint in 
$H\to \tau\mu$ decays would therefore falsify the (S)LH Models. On the contrary, if no LFV Higgs decays are confirmed at the LHC and recent LU breaking hints are refuted, these 
models with collective electroweak symmetry breaking will remain to be a promising alternative for the ultraviolet (UV) completion of the SM. Our aim in this paper is to make the above 
considerations precise, keeping in mind that current LHC analyses using LH models \cite{LHanalyses} do not show noticeable deviations.\\
In the next section we outline the main features of the SLH model, referring to our earlier work \cite{Lami:2016vrs} for an extended discussion. In section \ref{Results} we compute 
the leading contributions to the $H\to \ell\ell'$ decays within this model explaining our expansion in powers of the electroweak symmetry breaking scale ($v$) over that of the new particles 
of the extended model ($f$). We also confront the predictions of the SLH model for the considered decays to the current bounds, employing a parameter space of the SLH 
model which is consistent with current data and in section \ref{conclusions} we state our conclusions.\\

\section{The Simplest Little Higgs Model}
We reviewed in some detail the SLH model in our previous work \cite{Lami:2016vrs}. Here we will only recall its main features for the reader's convenience, but refer to 
\cite{Lami:2016vrs} and references therein for an expanded account.\\
The SLH model extends the SM electroweak gauge group $SU(2)_L\otimes U(1)_Y$ to $SU(3)_L\otimes U(1)_X$ in a minimal way. Correspondingly, the $SU(2)_L$ SM fermion doublets are enlarged to 
$SU(3)_L$ triplets and five extra weak gauge bosons are added (three neutral and two of opposite unit charges which, after diagonalization, give rise to heavy copies of the $W$ bosons that 
we call $W'^\pm$). $SU(3)$ invariant interactions are written so as to reproduce all the SM dynamics when restricted to the SM fields. Among the new particles, only the heavy quasi-Dirac 
neutrinos $N_k$ ($k=1,2,3$ is the family index), which allow the LFV transitions, and the $W'^\pm$ bosons play a role in our study. Within the SLH model, LFV is achieved because of the 
misalignment between the SM down-type lepton and the new heavy neutrino mass matrices.\\
The starting global symmetry of the model, $\left[SU(3)\otimes U(1)\right]_1\otimes\left[SU(3)\otimes U(1)\right]_2$, is spontaneously broken down to $\left[SU(2)\otimes U(1)\right]_1\otimes
\left[SU(2)\otimes U(1)\right]_2$ by the non-vanishing aligned vacuum expectation values (vevs) acquired by the two scalar triplets of the model \footnote{In simple group models (like SLH), 
where the SM gauge group emerges from the diagonal breaking of a larger simple group, at least two sigma-model multiplets are required \cite{SLH}.} (the scalar sector is a non-linear sigma 
model). These transform as $(3,1)$ and $(1,3)$ under $SU(3)_1\otimes SU(3)_2$, respectively, and include the SM Higgs doublets as well as new pseudo-Goldstone bosons (pGbs). Both vevs, 
$f_1$ and $f_2$, are of order TeV and determine the high-energy scale of the model, $f$, through $f_1^2+f_2^2=f^2$. The gauged diagonal subgroup, $SU(3)_L\otimes U(1)_X$, breaks down to the 
$SM$ electroweak gauge group via these scalar vacuum condensates. The gauge interactions of the model are fixed by gauge invariance and given in terms of the SM couplings.\\
The spontaneous symmetry breaking produces 5 pGbs for each scalar. Since one $SU(3)$ is weakly gauged, 5 pGbs give -through the Higgs mechanism- masses of $\mathcal{O}(f)$ to the new 
gauge bosons, while the 5 orthogonal combinations are pGbs (including the Higgs boson and, particularly, alleviating the hierarchy problem on its mass, $M_H\sim\mathcal{O}(v)$, due to 
the structure of the SLH model).
Since the mass of the new gauge bosons is proportional to the high scale, $f$: $M_{W'}=\frac{gf}{\sqrt{2}}\left(1+\mathcal{O}\left(\frac{v^2}{f^2}\right)\right)$, this suppresses their 
contribution at low energies. Light and the new heavy neutrinos do mix as regulated by the small parameter $\delta_\nu=\frac{-1}{\sqrt{2}\mathrm{tan}\beta}\frac{v}{f}$, where 
$\mathrm{tan}\beta=f_1/f_2$ is the ratio between both vevs, and heavy neutrino masses are also set by the large scale $f$. In our computation we will exploit the fact that the ratio 
$v/f$ is small ($\lesssim0.1$ \cite{smallratio}) to expand our amplitudes in it and keep only the leading contribution.\\

\section{Results}\label{Results}
\begin{figure}[t]\centering
\includegraphics[scale=0.7]{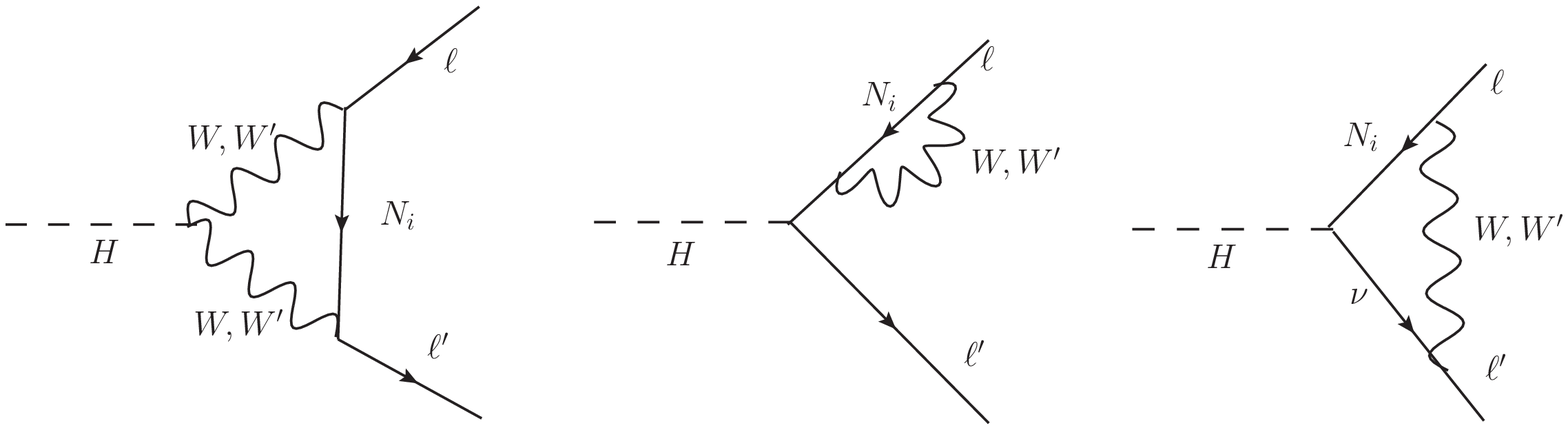}
\caption{\small }
\label{FeynmanDiagrams} Feynman diagrams for $H \to \ell \ell'$ decays in the SLH model.
\end{figure}

\begin{figure}[h!]
\vspace*{2.0cm}
\begin{center}
$ \begin{array}{cc}
\includegraphics[scale=0.33, angle=-90]{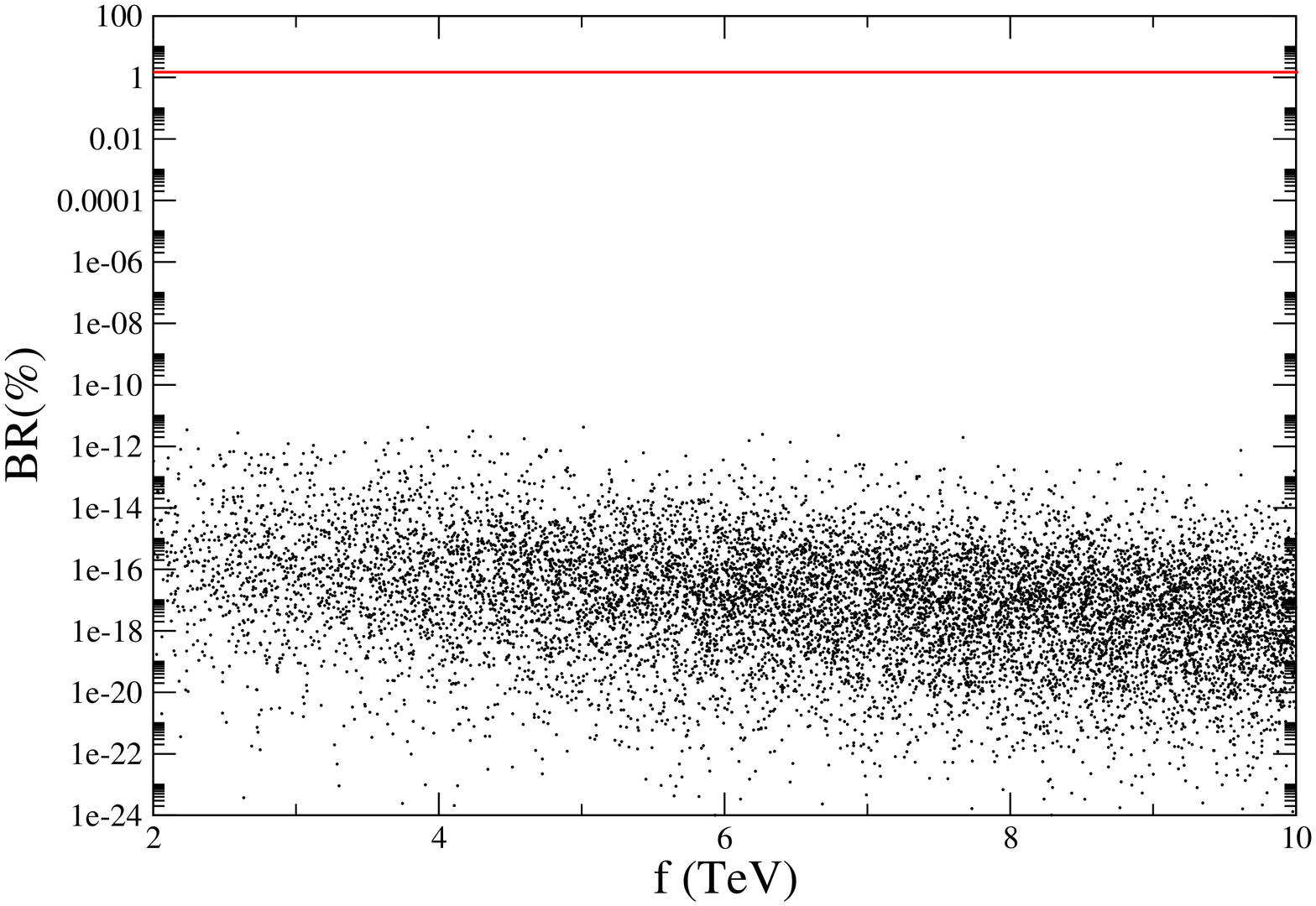}&
\includegraphics[scale=0.33, angle=-90]{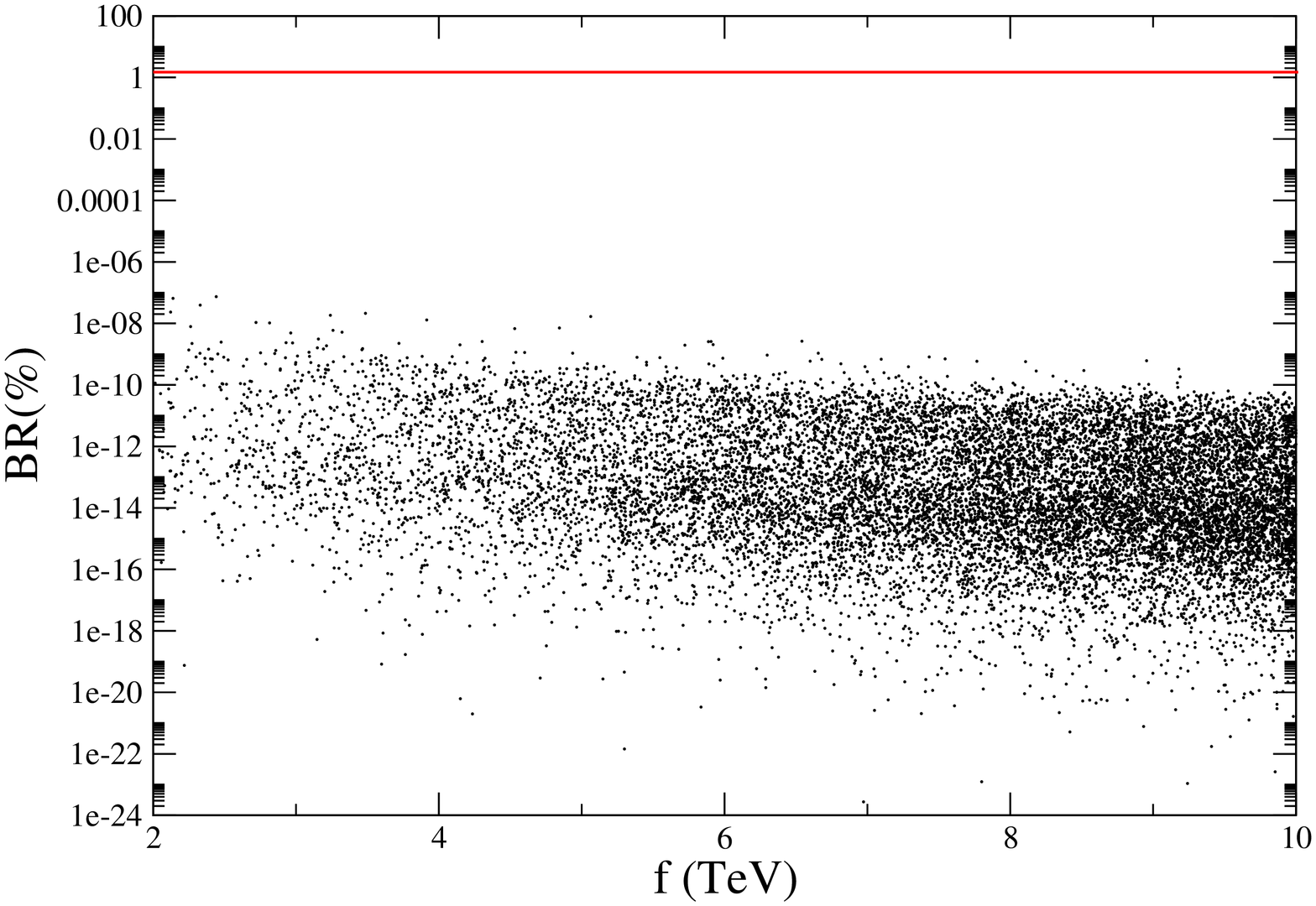}
 \end{array}$
 \end{center}
\caption{\label{plot1} Dependence of the scale of compositeness, $f$, of the branching ratio ($\%$) of the $H\to\tau\ell$ decays in the SLH model with two (left) and three (right) heavy neutrinos. 
The red line shows the 95$\%$ CL upper bound by CMS.}
\end{figure}

\begin{figure}[h!]
\vspace*{2.0cm}
\begin{center}
$ \begin{array}{cc}
\includegraphics[scale=0.33, angle=-90]{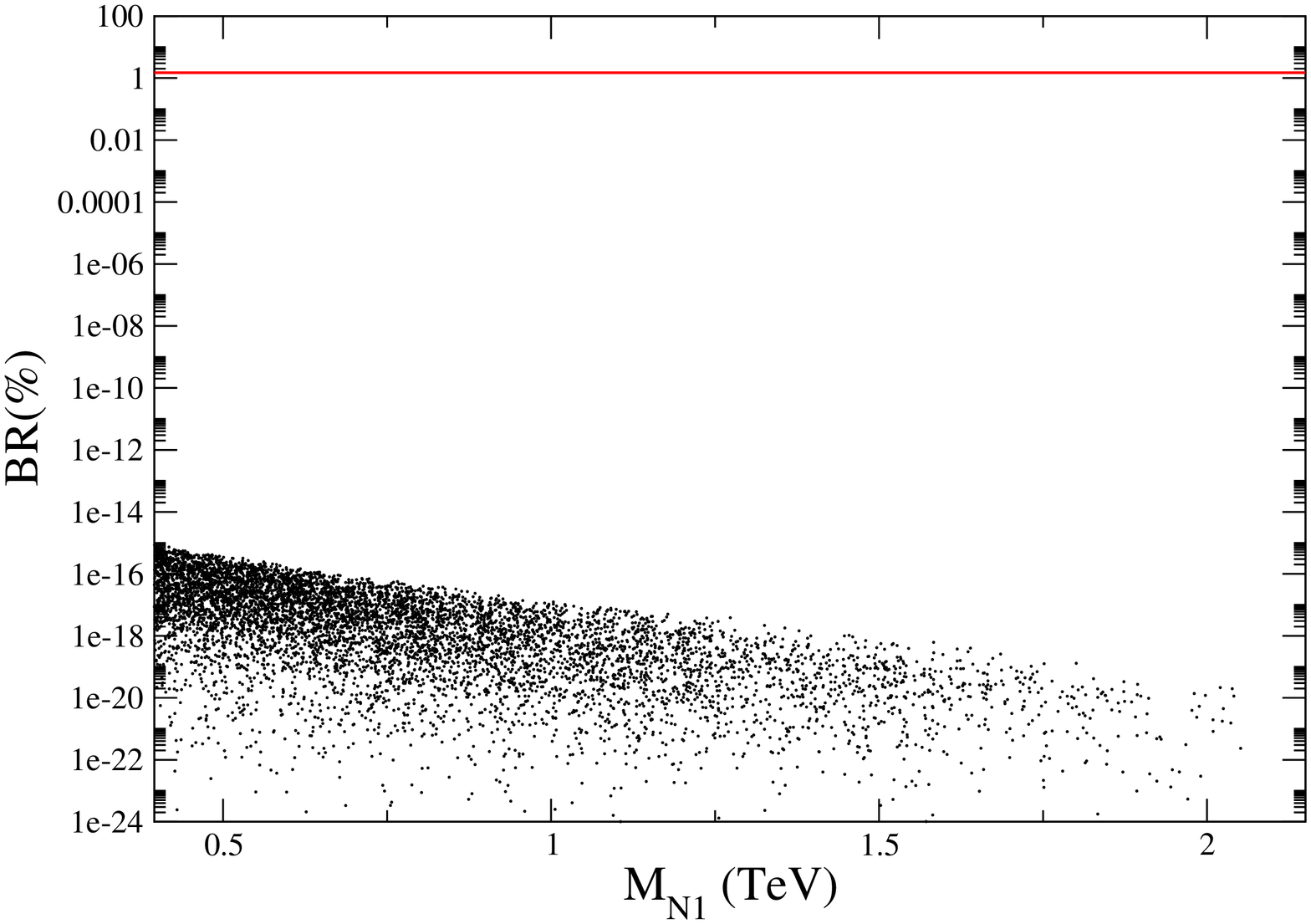}&
\includegraphics[scale=0.33, angle=-90]{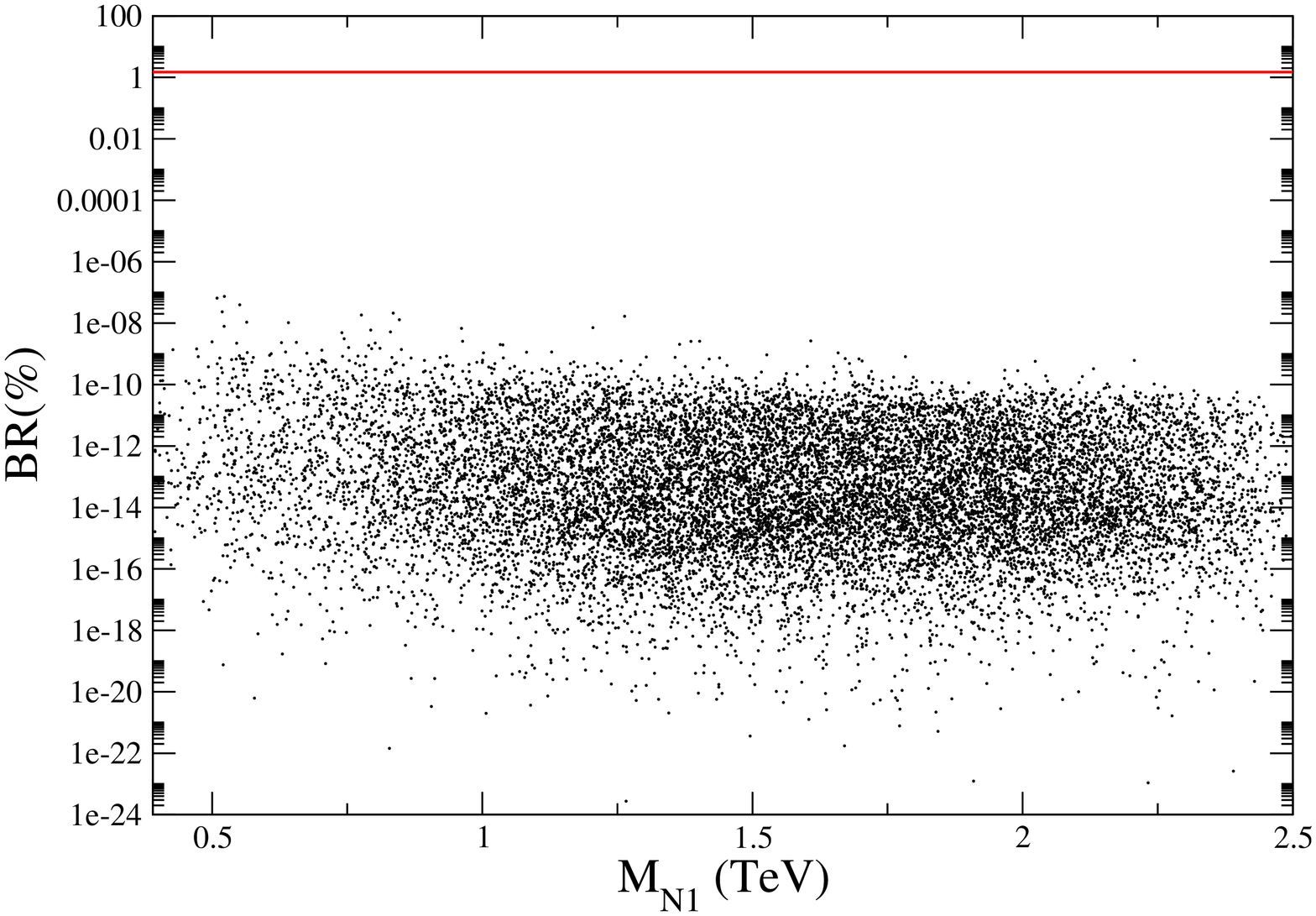}
 \end{array}$
 \end{center}
\caption{\label{plot2} Dependence on the lightest mass of the heavy neutrinos, $M_{N_1}$, of the branching ratio ($\%$) of the $H\to\tau\ell$ decays in the SLH model with two (left) and three (right) 
heavy neutrinos. The red line shows the 95$\%$ CL upper bound by CMS.}
\end{figure}

\begin{figure}[h!]
\vspace*{2.0cm}
\begin{center}
$ \begin{array}{cc}
\includegraphics[scale=0.33, angle=-90]{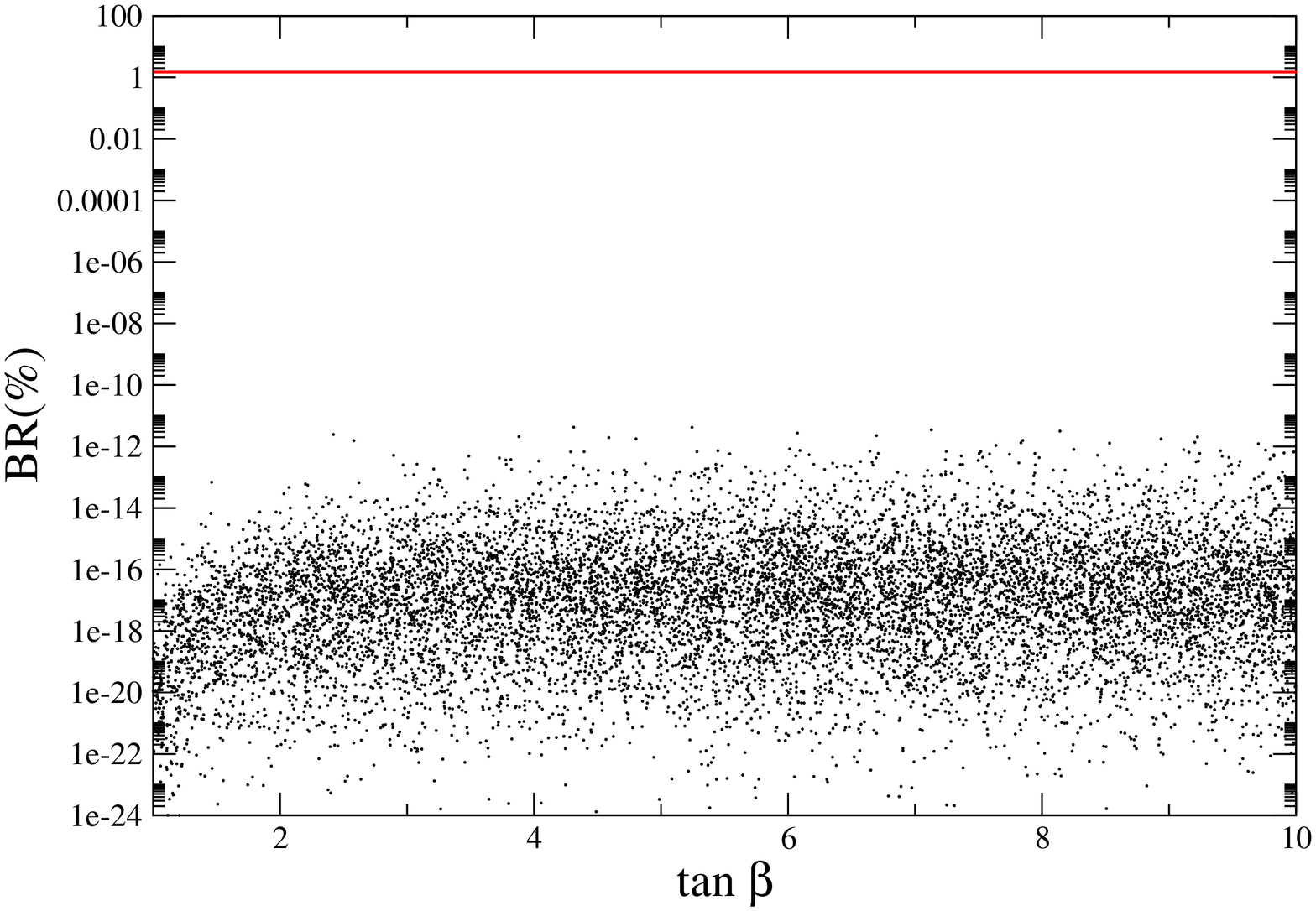}&
\includegraphics[scale=0.33, angle=-90]{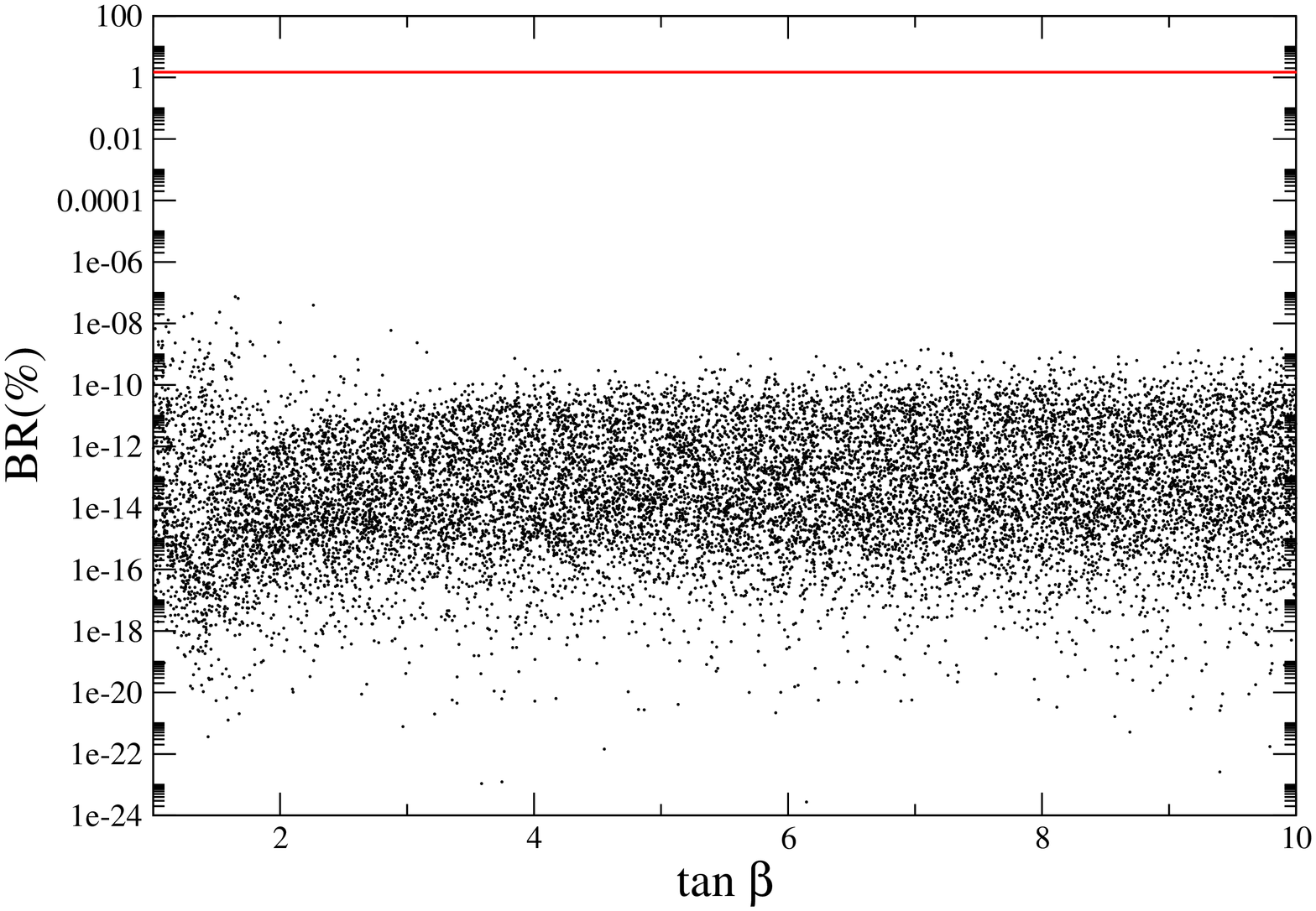}
 \end{array}$
 \end{center}
\caption{\label{plot3} Dependence on the ratio of the two vevs, $\tan \beta$, of the branching ratio ($\%$) of the $H\to\tau\ell$ decays in the SLH model with two (left) and three (right) heavy neutrinos. 
The red line shows the 95$\%$ CL upper bound by CMS.}
\end{figure}

\begin{figure}[h!]
\vspace*{2.0cm}
\begin{center}
$ \begin{array}{cc}
\includegraphics[scale=0.33, angle=-90]{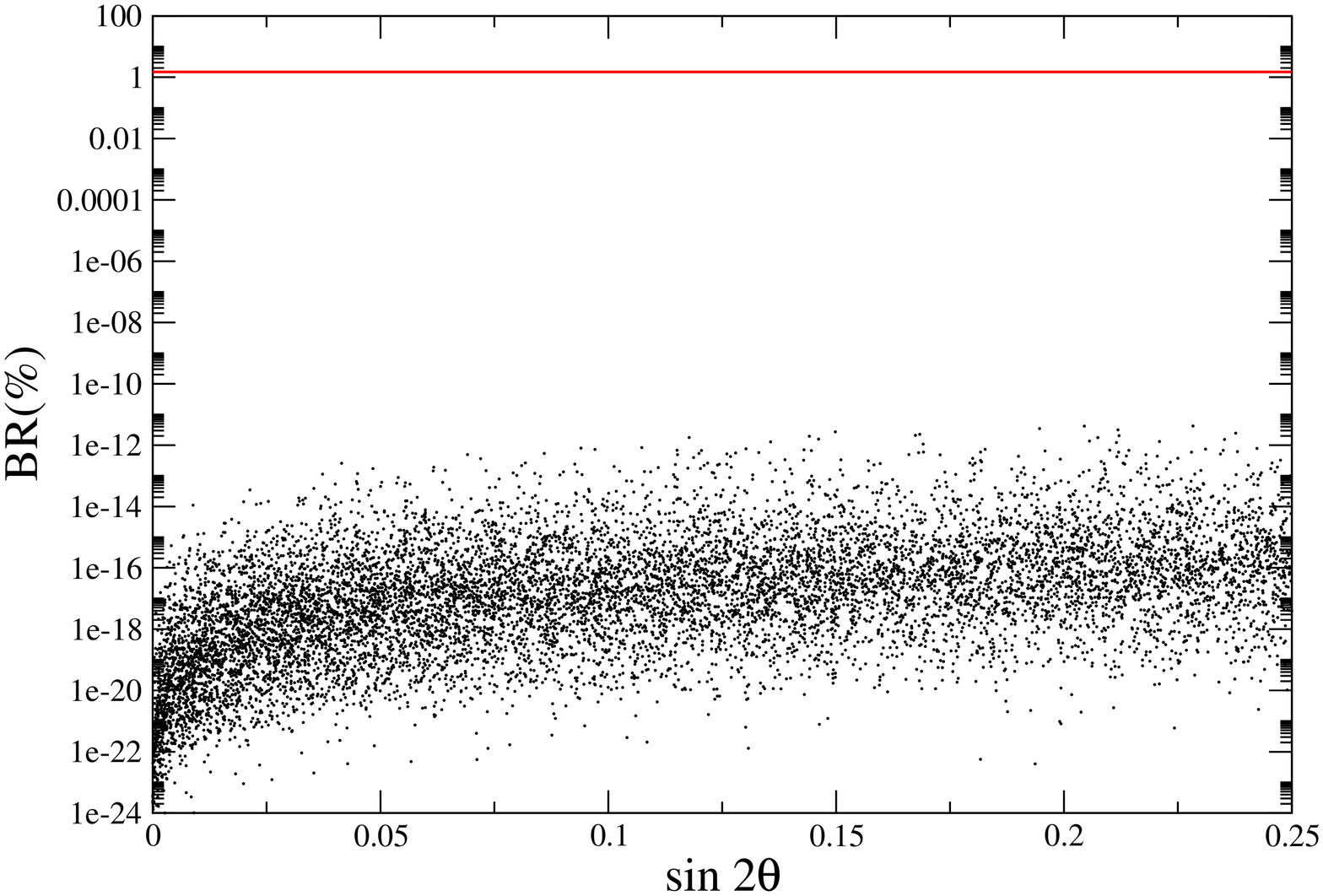}&
\includegraphics[scale=0.33, angle=-90]{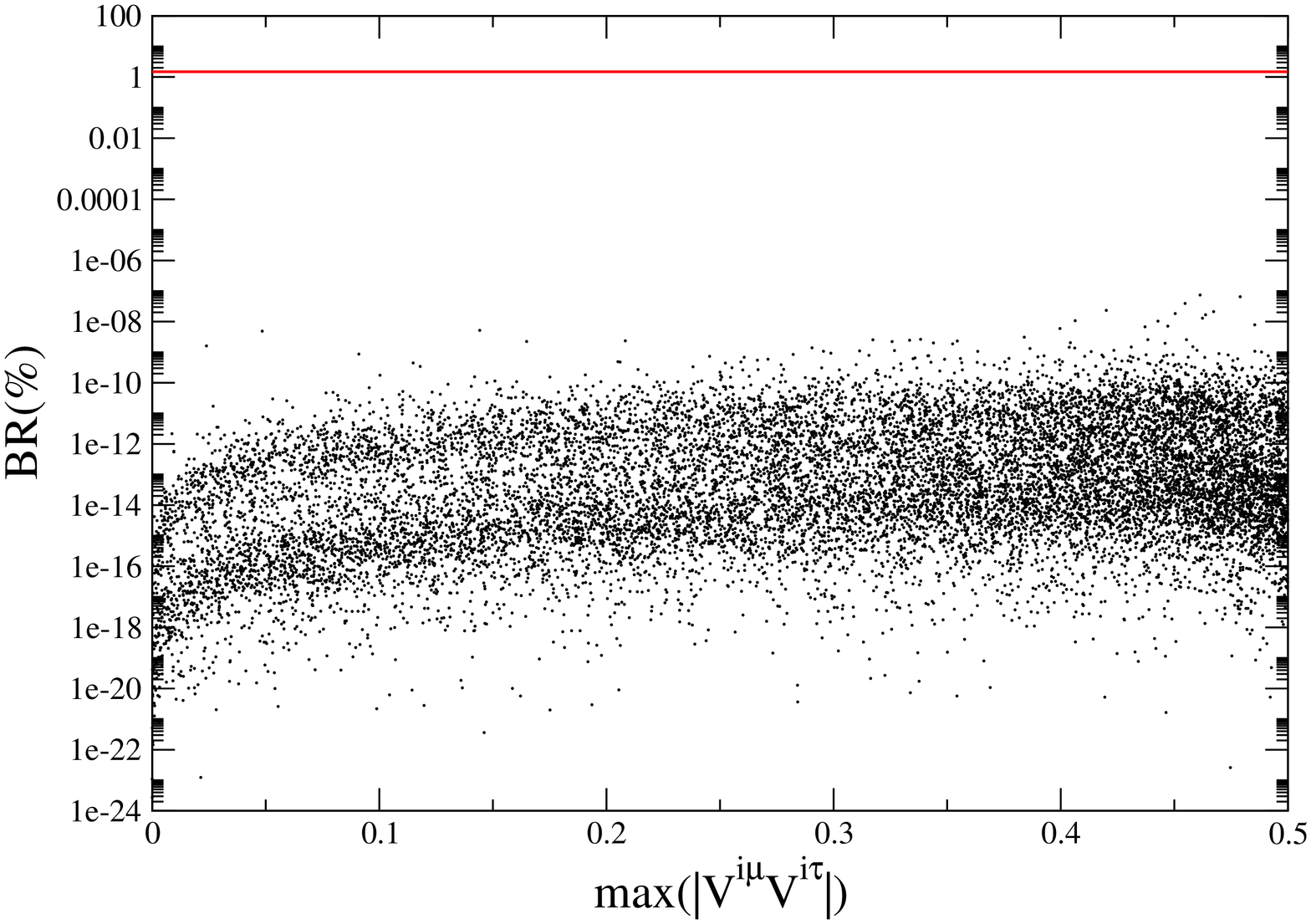}
 \end{array}$
 \end{center}
\caption{\label{plot4} Left plot: Dependence of the mixing angle between the two heavy leptons, $\sin 2 \theta$, of the branching ratio ($\%$) of the $H\to\tau\ell$ decays in the SLH model. The red line 
shows the 95$\%$ CL upper bound by CMS. Right plot: Analogous representation for the largest mixing among heavy neutrinos, $|V_\ell^{i\mu} V_\ell^{i\tau}|$.}
\end{figure}

LFV Higgs decays arise at one-loop level in the SLH model and are possible because the ``little`` heavy neutrinos $N_k$ couple to either charged lepton, $\ell_i$, irrespective of its flavor. 
In the considered $H\to \ell\ell'$ decays only the topologies sketched in Fig. \ref{FeynmanDiagrams} contribute at this order. Since the Higgs boson couples not only to a pair of 
$W^{(')}$ but also to $WW'$, the first topology gives rise to four different diagrams \footnote{We point out that exchanging $W\leftrightarrow W'$ in the diagrams built with the $HWW'$ vertex 
yields two different results, as can easily be shown.}. In the second topology, the loop mediating the LFV transition may as well be placed in the $\ell'$ leg, and in the last one the 
exchange $N_i\leftrightarrow\nu$ yields an inequivalent contribution. In all, this makes 12 different diagrams at this order working in the unitary gauge, where only physical degrees of 
freedom appear and the number of diagrams is reduced. Although each diagram diverges stronger in this gauge than in that of 't Hooft-Feynman, the individual divergences cancel, as they must, 
to ensure a finite result.\\
The contributions of self-energy type (second diagram) are proportional to $m_\ell$ and will thus be neglected for $\ell=e,\,\mu$. Along the computation we have neglected powers of the 
ratios of lepton masses over gauge boson ($W,W'$) and heavy neutrino ($N$) masses. For definiteness we include our results for the $H\to \tau\ell$ decay. There is an overall dependence on 
the heaviest final-state lepton mass, which shows that the decay rate $H\to \ell\ell'$ vanishes in the limit of massless decay products. Therefore, and in absence of a mechanism of LU 
violation in the SLH model, we will have $BR(H\to \tau\mu)=BR(H\to \tau e)=\frac{m_\tau^2}{m_\mu^2}BR(H\to \mu e)$, which suppresses the latter decay rate by a factor $\sim283$. Given this 
trivial proportionality, we will only be plotting $BR(H\to \tau\ell)$ in figs.~\ref{plot1}-\ref{plot4}.\\
Within this setting, there are two different mass scales in the problem: those of $\mathcal{O}(v)$ ($M_W$ and $M_H$) and those of $\mathcal{O}(f)$ ($M_N$ and $M_{W'}$). In Ref.~\cite{Lami:2016vrs} 
we used $\omega=\frac{M_W^2}{M_{W'}^2}\sim\frac{v^2}{f^2}<<1$ to characterize the ratio between two separated scales and $\chi_j=\frac{M_{N_j}^2}{M_{W'}^2}\sim\mathcal{O}(1)$ for that of two high scales. 
Since there were only three mass scales in the study of semileptonic LFV tau decays within this model, all mass ratios could be expressed in terms of $\omega$ and $\chi_j$ (unless there 
is a very strong hierarchy between the different heavy neutrino flavors). In the present study, there is $M_H$, as well. This entails the appearance of four small ratios between a 
light and a heavy particle mass: $\frac{M_H^2}{M_{N_j}^2}\sim\frac{M_W^2}{M_{N_j}^2}\sim\frac{M_H^2}{M_{W'}^2}\sim\frac{M_W^2}{M_{W'}^2}=\omega\sim\frac{v^2}{f^2}<<1$ (we recall that 
$\delta_\nu\sim\frac{v}{f}$ as well) and two involving particles with similar masses: $\frac{M_W^2}{M_H^2}\sim\frac{M_{N_j}^2}{M_{W'}^2}=\chi_j\sim\mathcal{O}(1)$.\\

Our analytical expressions are simplified in the limit of only two heavy neutrinos that we follow \cite{Lami:2016vrs}. We will however, consider also the case with three 
heavy neutrinos for completeness after eq.~(\ref{BR}). 
Let us discuss our choices for defining the parameter space of the model before presenting our expressions. In the numerical analysis we have stick to the choices argued in our previous work 
\cite{Lami:2016vrs}, in such a way that a considerable portion of the points generated randomly in the ranges fixed \textit{a priori} fulfils the constraints coming from 
$\mu\to e\gamma$, $\mu\to e e e$ and $\mu-e$ conversion in nuclei \cite{delAguila:2011wk} and also those on $\tau\to \mu\gamma$ \cite{Lami:2016vrs}.

We recall in the following the \textit{a priori} range of variation that we are allowing for the independent model parameters in our parameter space scan:
\begin{itemize}
 \item We have varied the scale of compositeness between 2 and 10 TeVs. Lower values are in tension with electroweak precision observables and larger figures enter the region where a UV 
 completion of the SLH model (that would become strongly coupled) starts to be expected \cite{smallratio}.
 \item The LFV processes are possible in the SLH model because of the presence of the heavy neutrinos. The dependence of the amplitude on their contribution is 
 \begin{equation} \label{eq:sumneu}
{\cal T} \, = \, \sum_{j} V_{\ell}^{j \mu *} V_{\ell}^{j \tau} \, A \left(\chi_j \right)\,.
\end{equation}
Assuming two families and one mixing angle, this can be written
\begin{equation} \label{eq:sumneu2}
{\cal T} \, = \, \frac{\sin 2\theta}{2}  \, \left[ A(\chi_1) - A(\chi_2) \right] \, . 
\end{equation}
This simplification will be used to write eq.(\ref{BR}) below. Particularly, since the terms with $\log(\omega)$ are independent on $\chi_j$, they do not contribute in this limit (see, however, eq.~(\ref{replacement3Ns})). 
This, by the way, prevents the appearance of a moderately large log, $\log(\chi_j/\omega)$, in the two heavy neutrino scenario. In the numerics, we will use the limits $0\leq \chi_1\leq0.25$ and 
$1.1\chi_1\leq \chi_2\leq 10\chi_1$ and $\sin 2 \theta \leq 0.25$, consistent with current data~\cite{delAguila:2011wk,heavyneutrinos}.
\item Finally, the ratio of the two vevs in the model, $\tan\beta$, is also a free parameter of the SLH model. We will take the range $ 1 < \tan \beta < 10$ for it, 
consistent with the known limits for the mixing between a 'little' and a light neutrino encoded in $\delta_\nu$, which should be $\lesssim 0.05$ \cite{delAguila:2011wk,heavyneutrinos}.
\end{itemize}

In these LFV Higgs decays there is at least a lepton which can be considered massless and, thus, with fixed helicity. Then, the matrix element for the $H\to \tau\ell$ decays can be 
written as

\begin{equation}\label{m.e.}
\mathcal{M}_{H}=-\frac{i v \alpha^2 \delta_\nu m_\tau\sum_j V_{\tau j} V_{\mu j}^* }{s_w^3 M_W^2}\left[O\log\chi_j-P(\chi_j)\right]\bar{u}(p') P_R u(q,m_\tau)\,,
\end{equation}
\\
where:

\begin{equation}
O=\frac{\delta_\nu}{16}\frac{M_H^2-13M_W^2}{M_W^2}
\end{equation}
\\
and

\begin{eqnarray}
P(\chi_j)&=&\mathrm{cot}\left(2\beta\right)\frac{M_H^2\left(3\chi_j^3-9\chi_j^2+8\chi_j-2\right)+M_W^2\chi_j\left(12\chi_j^2-23\chi_j+10\right)}{8M_{W'}M_W\chi_j\left(\chi_j-1\right)}
+\frac{\omega\left(2\chi_j^4+20\chi_j^3-74\chi_j^2+35\chi_j-1\right)}{48\left(\chi_j-1\right)^2}\nonumber\\
&+&\frac{\delta_\nu \chi_j}{24M_W^2\sin^2\left(2\beta\right)}\left[-5M_H^2+3\cos\left(4\beta\right)\left(4M_W^2-M_H^2\right)\right]\,.
\end{eqnarray}

In our computation we kept terms of subleading order ($v^3/f^3$) and check for accidental numerical enhancements of these before neglecting them. After checking its irrelevance, we 
omitted one such a term in $O$ and another one in $P(\chi_j)$.

The corresponding branching ratio is \footnote{The assumption of two heavy neutrinos has already been used.}

\begin{eqnarray}\label{BR}
BR\left(H\rightarrow\tau\mu\right)=\frac{(M_H^2-m_\tau^2)^2\alpha^4v^2\delta_\nu^2m_\tau^2}{16\pi M_H^3\Gamma_HM_W^4s_w^6}\left(\frac{\mathrm{sin}2\theta}{2}\right)^2\left[O
\log\left(\frac{\chi_1}{\chi_2}\right)+P(\chi_2)-P(\chi_1)\right]^2\,.
\end{eqnarray}
\\

We have performed a scan of the parameter space limited by the above \textit{a priori} restrictions verifying that the constraints from low-energy 
processes are respected and plotted $BR(H\to\tau\ell)$ ($\%$) as a function of one parameter in turn ($f$, $M_{N_1}$, $\tan \beta$ and $\sin 2 \theta$) in the left pannel 
of figs. \ref{plot1}-\ref{plot4}. Roughly $23\%$ of the $5\cdot10^4$ randomly generated points satisfied the low-energy restrictions.

Since these branching fractions are very suppressed, we have considered next the case with three heavy neutrinos, $N_k$, expecting that the increase of degrees of freedom allows to satisfy the low-energy contraints on LFV processes 
and, at the same time, yield $H\to\ell\ell'$ with larger decay rates. In this general case we have considered the PDG \cite{pdg} parametrization for the mixing between charged leptons and heavy neutrinos assuming only a vanishing CP violating phase 
for simplicity. 
Therefore

\begin{equation}
 V_{\ell}^{if}\,=\,\left(
 \begin{array}{ccc}
  c_{12}c_{13} & s_{12}c_{13} & s_{13} \\
  -s_{12}c_{23}-c_{12}s_{23}s_{13} & c_{12}c_{23}-s_{12}s_{23}s_{13} & s_{23}c_{13} \\
  s_{12}s_{23}-c_{12}c_{23}s_{13} & -c_{12}s_{23}-s_{12}c_{23}s_{13} & c_{23}c_{13}
 \end{array}
 \right)\,,
\end{equation}
where we have employed the usual notation $c_{ij}\,\equiv$ cos$\theta_{ij}$, $s_{ij}\,\equiv$ sin$\theta_{ij}$, $i\,=\,1,\,2,\,3$ and $f\,=\,e,\,\mu,\,\tau$. In this case the replacement 
needed in eq.~(\ref{BR}) is
\begin{equation} \label{replacement3Ns}
 \left(\frac{\mathrm{sin}2\theta}{2}\right)^2\left[O \log\left(\frac{\chi_1}{\chi_2}\right)+P(\chi_2)-P(\chi_1)\right]^2 \to \sum_{i=1}^3 V_\ell^{i\mu}V_\ell^{i\tau}\left[O\log\left(\frac{\chi_i}{\omega}
 \right)+P(\chi_i)\right]^2\,,
\end{equation}
where the assumed CP conservation of $V_{\ell}^{if}$ was used.

We have 
verified that restricting the maximum values of $|V_\ell^{i f} V_\ell^{i f^\prime}|$ to reasonable upper limits does not yield additional constraints than requiring the fulfilment of the 
experimental bounds on low-energy LFV processes. Consequently, we have scanned over $-1\leq s_{ij}\leq 1$ ensuring the low-energy restrictions. Constraints in the case with three heavy neutrinos are milder than in the previous case where two 
heavy neutral leptons were considered \cite{heavyneutrinos}. We take advantage of that to slightly modify our choices of their parameters aiming to increase the predicted $H\to\ell\ell'$ branching ratios. For the masses of the three heavy neutrinos 
we have followed the educated guess $0.1\leq\chi_1\leq0.25$, $1.1\chi_1\leq\chi_2\leq100\chi_1$ and $1.1\chi_2\leq\chi_3\leq100\chi_2$ (we recall that $\chi_i$ depends quadratically on the $N_i$ mass), while $f$ and tan$(\beta)$ have been varied 
in the same way as in the two heavy neutrino scenario. In this way we have obtained the plots depicted on the right pannels of figures \ref{plot1}-\ref{plot4}. The comparison of the left and right plots of these figures shows that, indeed, allowing 
for an extra heavy neutrino gives us enough freedom to increase the predicted decay rates by some four orders of magnitude. However, these still remain well below the CMS hint on $H\to\tau\mu$ and the detectability level at LHC.

\begin{figure}[h!]
\vspace*{2.0cm}
\includegraphics[scale=0.6, angle=-90]{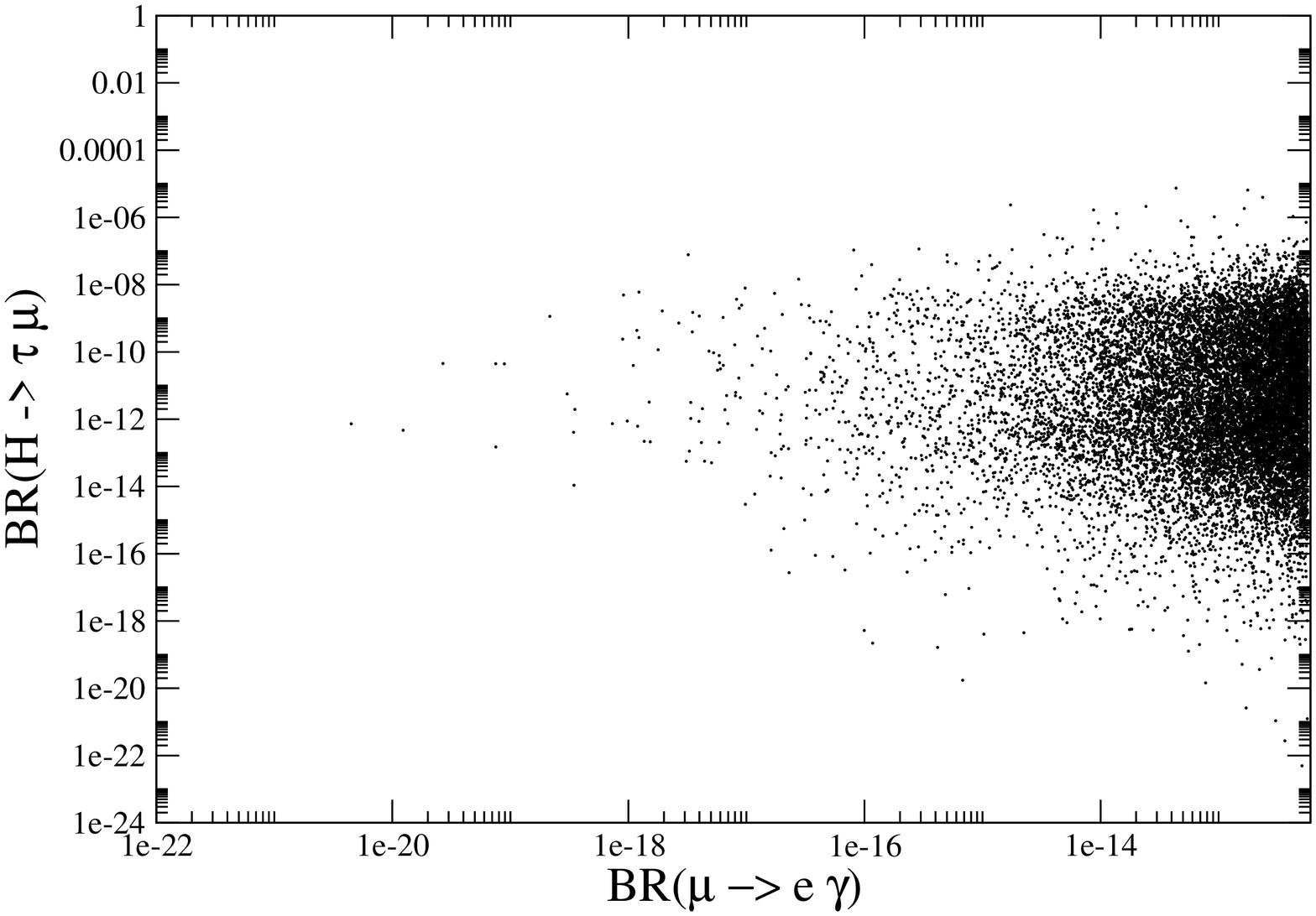}
\caption{\label{plot5} The correlation between the $H\to\tau\ell$ and $\mu\to e \gamma$ decays is illustrated within the SLH model in the case with three heavy neutrinos, $N_k$. The 
x-axis is cut at the current upper limit (UL) at 90$\%$ C.L. of $BR(\mu\to e\gamma)$, $5.7\cdot10^{-13}$~\cite{pdg}.}
\end{figure}

The general trend is that the SLH model produces $H\to\tau\ell$ decay widths which are at least six orders of magnitude smaller than the $BR\sim\mathcal{O}(\%)$ hinted 
by CMS (see also figure \ref{plot5}). A similarly strong suppression of $BR(H\to\tau\ell)$ is also found in a recent analysis within the Little Higgs Model with T-parity including 
constraints from other charged lepton flavor violating processes~\cite{Yang:2016hrh}. Finally, in figure \ref{plot5} we show that there is basically no correlation between $BR(H\to\tau\ell)$ and 
$BR(\mu\to e\gamma)$ (the most restrictive low-energy search) in the case with three heavy neutrinos. There are not sizable correlations between the pairs $BR(H\to\tau\ell)$, $BR(\tau\to\ell\gamma)$ and $BR(H\to\mu e)$, $BR(\mu\to e\gamma)$ either.

As expected, if the SLH model is to satisfy the bounds on $H\to\mu e$ set by $\mu\to e\gamma$ ($BR(H\to \mu e)\lesssim10^{-8}$ \cite{LowE}) 
-as it does-, it must fall way too short to 
explain the CMS hint, as a consequence of its LU. It must be pointed out, however, that we restricted the Yukawa interactions in the lepton sector up to operators of dimension 5. Given 
our ignorance of the flavor structure of the theory at its cut-off, it could be possible that the contribution of higher-dimensional operators (see, e.g. Harnik \textit{et. al.} in 
\cite{LowE}) could change the results we presented. Finally, we point out that very mild variations are appreciated in figs. \ref{plot1}-\ref{plot4} with respect to the independent variables. Decay 
probabilities are slightly larger for smaller $f$ and $M_{N_1}$ and for larger $\tan\beta$ and $\sin2\theta$ (max$|V_\ell^{i\mu} V_\ell^{i\tau}|$).\\

\section{Conclusions}\label{conclusions}
Flavor violation has shown up in the quark and neutrino sectors. Although extremely suppressed in the SM extended with right-handed neutrinos, it may appear at measurable rates in several 
well-motivated new physics models. On the other hand, the discovery of the Higgs boson at the LHC has brought a new scenario to search for LFV in the decays of this scalar. An elegant 
solution to the hierarchy problem on the Higgs mass is provided by the LH models. Here, we have considered the SLH model (one of the simplest realizations of these ideas) against the 
ATLAS and CMS limits on $BR(H\to\tau\mu)$, of order percent. Given the LU of the SLH model, it is not surprising that the model cannot simultaneously account for the tiny rate at which 
the $H\to\mu e$ decays must proceed ($\lesssim10^{-8}$) and also for a measurable signal at LHC. We have found that a $BR(H\to\mu e)$ as low as $10^{-10}$ is obtained naturally (even allowing for three 
heavy neutrinos, which increases the predicted decay rates with respect to the scenario with two heavy neutral leptons), with $BR(H\to\tau\ell)$ only enhanced by a factor of order $300$. Thus, the confirmation of the CMS 
hint would disfavor the SLH model, as it will do a measurement at Belle-II of semileptonic LFV tau decays \cite{Lami:2016vrs}.

\vspace*{0.25cm}
\section*{Acknowledgements}
We both thank Jorge Portol\'es for drawing our interest to charged lepton flavor violation and for valuable suggestions on this study. This research has been supported in part by the 
Spanish Government, Generalitat Valenciana and ERDF funds from the EU Commission [grants FPA2011-23778, FPA2014-53631-C2-1-P, PROMETEOII/2013/007, SEV-2014-0398]. A. L. also acknowledges 
Generalitat Valenciana for a Grisol\'{\i}a scholarship. P. R. benefited from the financial support of projects 296 ('Fronteras de la Ciencia'), 236394, 250628 ('Ciencia B\'asica') and 
SNI (Conacyt, Mexico), and from the hospitality of IFIC, where part of this work was done. P. R. is also indebted for a useful discussion on this topic with Juanjo Sanz Cillero.

\end{document}